\documentclass[twocolumn,aps,prl,amssymb,amsmath,twoside,showpacs]{revtex4}
\usepackage{epic,eepic}

\newcommand{\xd}{\mathrm{d}}
\newcommand{\xD}{\mathcal{D}}
\newcommand{\im}{\mathrm{i}}
\newcommand{\varphif}{\check{\varphi}}
\newcommand{\xhat}{\hat{x}}
\newcommand{\that}{\hat{t}}
\newcommand{\omegahat}{\hat{\omega}}
\newcommand{\pop}{\kappa}

\begin{document}
\title{States on timelike hypersurfaces in quantum field theory}
\author{Robert Oeckl}\email{robert@matmor.unam.mx}
\affiliation{Instituto de Matem\'aticas, UNAM Campus Morelia,
C.P. 58190, Morelia, Michoac\'an, Mexico}
\date{30 May 2005}
\pacs{11.10.-z, 03.70.+k}

\begin{abstract}
We investigate the possibility of defining states on timelike
hypersurfaces in quantum field theory.
To this end we consider hyperplanes in the real massive
Klein-Gordon theory using the Schr\"odinger
representation. We find a well defined vacuum wave functional, existing
on any hyperplane, with the remarkable property that it changes smoothly
even under Euclidean rotation through the light-cone. Moreover, particles
on timelike hyperplanes exist and
occur in two variants, incoming and outgoing,
distinguished by the sign of the energy. Multi-particle wave
functionals
take a form similar to those on spacelike hypersurfaces. The role of
unitarity and the inner product is discussed.
\end{abstract}

\maketitle

Traditionally, Hilbert spaces of states in quantum field theory
are associated with spacelike hypersurfaces. This is rooted
in quantization prescriptions relying on the initial value problem,
i.e., a correspondence between solutions of the
equations of motion and initial data on a spacelike hypersurface.
Furthermore, it may seem that causality requires hypersurfaces
carrying states to be
spacelike for a probability interpretation via the
inner product to make sense.

On the other hand, quantum field theory shows properties suggesting
that this restriction to spacelike hypersurfaces is
artificial. In particular, the crossing symmetry of transition
amplitudes may be taken as a strong hint that state spaces associated
with more general hypersurfaces should make sense
\cite{Oe:catandclock}. One may push this line of
reasoning further to arrive at an axiomatic formulation of quantum
field theory in the spirit of topological quantum
field theory \cite{Oe:boundary}. This view may also be motivated from
quantum gravity \cite{Oe:catandclock,Oe:boundary,CDORT:vacuum}.

We limit ourselves here to the investigation of states on
hypersurfaces that are hyperplanes and consider the real massive
Klein-Gordon theory in Minkowski space.
Since we need to be free from the constraints of a canonical approach
we employ the Schr\"odinger
representation of quantum field theory
\cite{Sym:schroedinger,Lue:schroedinger,Jak:schroedinger} together
with Feynman's path integral \cite{Fey:stnrqm}.

\section{Schr\"odinger representation}

The classical massive Klein-Gordon theory is defined by the wave
equation of motion $(\partial_0^2-\sum_{i\ge 1}
\partial_i^2+m^2)\phi=0$ for 
a real scalar field $\phi(x)$.
States in the quantum theory are wave functionals, i.e., functions on
field configurations at a given time. The Schr\"odinger equation is
formally solved by Feynman's path integral. Thus,
the time-evolution from a state $\Psi$ at time $t$
to a state $\Psi'$ at time $t'$ is expressible as
\begin{equation}
 \Psi'(\varphi') =
 \int \xD \varphi\, \Psi(\varphi)
 \int_{\phi|_{t}=\varphi, \phi|_{t'}=\varphi'} \xD \phi\,
 e^{\im S(\phi)} .
\label{eq:tevol}
\end{equation}
The outer integral is over field configurations $\varphi$
at time $t$. The inner integral, called \emph{field propagator} and
denoted by $Z(\varphi,\varphi')$, is over
space-time field configurations $\phi$ in the time interval
$[t,t']$ which match the configurations $\varphi$ and $\varphi'$.
Since the action $S$ is quadratic, the stationary phase
approximation to this integral is exact.
Using this fact, rewriting the action as a boundary integral and
using a decomposition of classical solutions into positive and
negative energy components yields the propagator
\begin{equation}
 Z(\varphi,\varphi')=N \exp\left(-\frac{1}{2}\int\xd^3 x
 \begin{pmatrix}\varphi & \varphi' \end{pmatrix} W
 \begin{pmatrix}\varphi \\ \varphi' \end{pmatrix}\right) .
\label{eq:proptbdy}
\end{equation}
The operator-valued matrix $W$ is given by
\[
 W=\frac{-\im\omega}{\sin\omega\Delta}
 \begin{pmatrix}\cos\omega\Delta & -1 \\
  -1 & \cos\omega\Delta \end{pmatrix} ,
\]
where $\Delta=t'-t$ and
$\omega=\sqrt{-\sum_{i\ge 1} \partial_i^2+m^2}$.

The vacuum state is characterized by its invariance
under time-evolution. We make the ansatz
\begin{equation}
 \Psi_0(\varphi)=C \exp\left(-\frac{1}{2}\int \xd^3 x\, \varphi(x)
(A \varphi)(x)\right)
\label{eq:vacansatz}
\end{equation}
for an unknown operator $A$.
Evolving the state using the propagator (\ref{eq:proptbdy}) we find
that invariance is
equivalent to the equation $A^2=\omega^2$. We make the conventional
choice $A=\omega$.
A one-particle state of momentum $p$ is given by
\begin{equation}
\Psi_p(\varphi)=\varphif(p) \Psi_0(\varphi) ,
\label{eq:pstate}
\end{equation}
where $\varphif$ is the Fourier transform
$\varphif(p)=2E \int \xd^3 x\, e^{\im p x} \varphi(x)$.

\section{Vacuum on boosted hyperplanes}

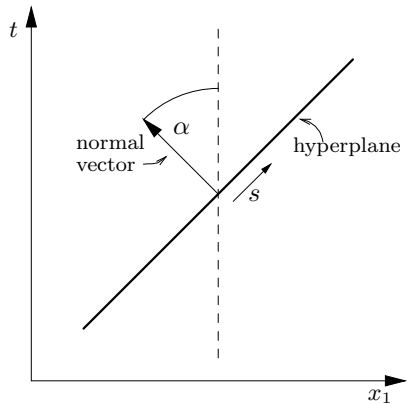
\begin{figure}
\setlength{\unitlength}{0.00087489in}
\begingroup\makeatletter\ifx\SetFigFont\undefined%
\gdef\SetFigFont#1#2#3#4#5{%
  \reset@font\fontsize{#1}{#2pt}%
  \fontfamily{#3}\fontseries{#4}\fontshape{#5}%
  \selectfont}%
\fi\endgroup%
{\renewcommand{\dashlinestretch}{30}
\begin{picture}(2483,2443)(0,-10)
\put(1260.620,1291.571){\arc{1272.862}{3.9257}{4.7098}}
\path(135,166)(2385,166)
\blacken\path(2265.000,136.000)(2385.000,166.000)(2265.000,196.000)(2265.000,136.000)
\path(135,166)(135,2416)
\blacken\path(165.000,2296.000)(135.000,2416.000)(105.000,2296.000)(165.000,2296.000)
\dashline{60.000}(1260,301)(1260,2281)
\thicklines
\path(450,481)(2070,2101)
\thinlines
\path(1260,1291)(810,1741)
\blacken\path(916.066,1677.360)(810.000,1741.000)(873.640,1634.934)(916.066,1677.360)
\path(1350,1246)(1575,1471)
\blacken\path(1543.180,1417.967)(1575.000,1471.000)(1521.967,1439.180)(1543.180,1417.967)
\path(1890,1606)(1889,1609)(1888,1614)
	(1886,1623)(1883,1633)(1879,1644)
	(1874,1656)(1867,1669)(1857,1682)
	(1845,1696)(1831,1708)(1818,1718)
	(1805,1725)(1793,1730)(1782,1734)(1755,1741)
\path(1816.844,1740.462)(1755.000,1741.000)(1809.315,1711.422)
\path(810,1471)(811,1471)(818,1471)
	(832,1472)(848,1472)(863,1473)
	(875,1475)(884,1476)(893,1478)
	(900,1481)(907,1485)(915,1491)
	(924,1498)(945,1516)
\path(909.206,1465.564)(945.000,1516.000)(889.683,1488.341)
\put(0,2236){\makebox(0,0)[lb]{{\SetFigFont{9}{10.8}{\rmdefault}{\mddefault}{\updefault}$t$}}}
\put(990,1651){\makebox(0,0)[lb]{{\SetFigFont{10}{12.0}{\rmdefault}{\mddefault}{\updefault}$\alpha$}}}
\put(405,1561){\makebox(0,0)[lb]{{\SetFigFont{8}{9.6}{\rmdefault}{\mddefault}{\updefault}normal}}}
\put(405,1426){\makebox(0,0)[lb]{{\SetFigFont{8}{9.6}{\rmdefault}{\mddefault}{\updefault}vector}}}
\put(1710,1516){\makebox(0,0)[lb]{{\SetFigFont{8}{9.6}{\rmdefault}{\mddefault}{\updefault}hyperplane}}}
\put(2160,31){\makebox(0,0)[lb]{{\SetFigFont{9}{10.8}{\rmdefault}{\mddefault}{\updefault}$x_1$}}}
\put(1440,1246){\makebox(0,0)[lb]{{\SetFigFont{10}{12.0}{\rmdefault}{\mddefault}{\updefault}$s$}}}
\end{picture}
}
\caption{Position of the hyperplane in $(t,x_1)$-coordinates. The
  $s$-coordinate runs along the hyperplane.}
\label{fig:hyperplane}
\end{figure}

We shall now be interested in states on spacelike hyperplanes that
are not normal to the time axis. Since the effect of spatial rotations
is essentially trivial it suffices to single out the $x_1$-coordinate
and consider hyperplanes whose normal
vector lies in the plane spanned by $x_1$ and $t$, see
Figure~\ref{fig:hyperplane}. We shall express
everything in terms of \emph{Euclidean} coordinates and angles with
respect to the original coordinate system. Thus, call $\alpha$ the
angle between the hyperplane and the time axis. Call $s$ the
rotated version of the $x_1$-coordinate.
Thus, within the hyperplane, $x_1=s \cos\alpha$ and
$t=s \sin\alpha$. The hyperplane is spacelike
if $\alpha < \pi/4$.

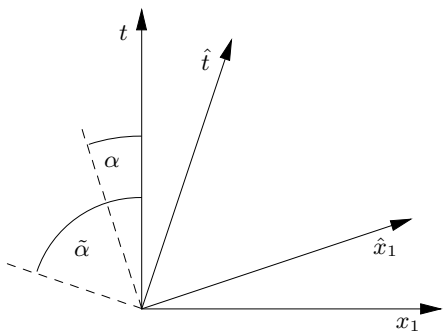
\begin{figure}
\setlength{\unitlength}{0.00087489in}
\begingroup\makeatletter\ifx\SetFigFont\undefined%
\gdef\SetFigFont#1#2#3#4#5{%
  \reset@font\fontsize{#1}{#2pt}%
  \fontfamily{#3}\fontseries{#4}\fontshape{#5}%
  \selectfont}%
\fi\endgroup%
{\renewcommand{\dashlinestretch}{30}
\begin{picture}(2675,1993)(0,-10)
\put(812.448,179.574){\arc{1310.964}{3.4700}{4.7254}}
\put(811.891,232.789){\arc{1944.507}{4.3934}{4.7218}}
\path(822,166)(822,1966)
\blacken\path(852.000,1846.000)(822.000,1966.000)(792.000,1846.000)(852.000,1846.000)
\path(822,166)(2622,166)
\blacken\path(2502.000,136.000)(2622.000,166.000)(2502.000,196.000)(2502.000,136.000)
\path(822,166)(1362,1786)
\blacken\path(1352.513,1662.671)(1362.000,1786.000)(1295.592,1681.645)(1352.513,1662.671)
\path(822,166)(2442,706)
\blacken\path(2337.645,639.592)(2442.000,706.000)(2318.671,696.513)(2337.645,639.592)
\dashline{60.000}(822,166)(462,1246)
\dashline{60.000}(822,166)(12,436)
\put(2352,31){\makebox(0,0)[lb]{{\SetFigFont{9}{10.8}{\rmdefault}{\mddefault}{\updefault}$x_1$}}}
\put(597,1021){\makebox(0,0)[lb]{{\SetFigFont{9}{10.8}{\rmdefault}{\mddefault}{\updefault}$\alpha$}}}
\put(417,481){\makebox(0,0)[lb]{{\SetFigFont{9}{10.8}{\rmdefault}{\mddefault}{\updefault}$\tilde{\alpha}$}}}
\put(2217,481){\makebox(0,0)[lb]{{\SetFigFont{9}{10.8}{\rmdefault}{\mddefault}{\updefault}$\xhat_1$}}}
\put(687,1786){\makebox(0,0)[lb]{{\SetFigFont{9}{10.8}{\rmdefault}{\mddefault}{\updefault}$t$}}}
\put(1182,1606){\makebox(0,0)[lb]{{\SetFigFont{9}{10.8}{\rmdefault}{\mddefault}{\updefault}$\that$}}}
\end{picture}
}
\caption{Original $(t,x_1)$ and boosted $(\that,\xhat_1)$
  coordinates.}
\label{fig:lorentz}
\end{figure}

Since the theory is manifestly
Lorentz covariant we can produce states on these spacelike
hyperplanes from those on the standard hyperplane by Lorentz boosts.
Consider a boost in the $(t,x_1)$-plane with parameter
$\gamma$. The boosted coordinates $\that,\xhat_1$ are given in terms
of the original ones by
\begin{equation}
\begin{pmatrix} \that \\ \xhat_1 \end{pmatrix}
 = \begin{pmatrix} \cosh\gamma & -\sinh\gamma \\
 -\sinh\gamma & \cosh\gamma \end{pmatrix}
\begin{pmatrix} t \\ x_1 \end{pmatrix} ,
\label{eq:boost}
\end{equation}
see Figure~\ref{fig:lorentz}.
We shall consider the hyperplane spanned by $(\xhat_1,x_2,x_3)$. To bring
this into accordance with the previous picture, the angles $\alpha$ in
Figures~\ref{fig:hyperplane} and \ref{fig:lorentz} have to agree. The
relation between $\xhat_1$ and $s$ is then given by the length
contraction factor $\rho=1/\sqrt{\cosh 2\gamma}$
via $\xhat_1=\rho s$. The Euclidean angle $\alpha$ is related to $\rho$
via $\cos 2\alpha=\rho^2$.

Consider now the vacuum state on the hyperplane spanned by
$(s,x_2,x_3)$. In terms of the boosted coordinates $(\xhat_1,x_2,x_3)$, it
takes the usual form given by (\ref{eq:vacansatz}) with $A=\omegahat$
(we use a hat to indicate that the coordinate $\xhat_1$ instead of
$x_1$ appears in $\omega$ ).
Transforming the expression to $(s,x_2,x_3)$-coordinates affects both
the operator $\omegahat$ as well as the integral measure $\xd \xhat_1$.
Explicitly, $\xd \xhat_1=\rho\, \xd s$ and
$\omegahat=\sqrt{-\rho^{-2}\partial_s^2-\sum_{i\ge 2}\partial_i^2+m^2}$,
where $\partial_s$ is the derivative in the
$s$-coordinate. Abbreviating $(x_2,x_3)$ collectively by $\tilde{x}$, we
can write the resulting expression for the vacuum as
\begin{equation}
 \Psi_0(\varphi)=C \exp\left(-\frac{1}{2}\int \xd s\, \xd^2 \tilde{x}\,
 \varphi(s,\tilde{x}) (\tau \varphi)(s,\tilde{x})\right),
\label{eq:vacgeneral}
\end{equation}
where
$\tau = \sqrt{-\partial_s^2+\cos 2\alpha\,
(-\sum_{i\ge 2}\partial_i^2+m^2)}$.

One might be surprised that the resulting expression takes a simple
form in terms of \emph{Euclidean} coordinates and angles.
What is more, the vacuum
functional appears to behave smoothly in the limit $\alpha\to \pi/4$
and even beyond the light-cone. Is it meaningful there?

\section{Vacuum on timelike hyperplanes}

Note that we could have used an alternative route to obtain a general
expression for the vacuum functional. Namely, for a given
hyperplane
we could have worked out the field propagator to a parallel hyperplane
and then imposed invariance under the propagation using the analogue of
the ansatz (\ref{eq:vacansatz}). The result would have been
identical to (\ref{eq:vacgeneral}) due to
the Lorentz covariance of our setup.
On the other hand, this suggests a characterization of the
vacuum beyond the spacelike case as being invariant under
propagation to a parallel hyperplane.

As for spacelike hyperplanes it suffices to start with a given one
and obtain any other one through a Lorentz transformation. Thus, we
single out the hyperplane spanned by the coordinates $(t,x_2,x_3)$.
We wish to define states in analogy with the spacelike case as
functions on field configurations on the hyperplane.
From a quantization point of view the configuration space should be
``half'' of the phase-space, which in turn corresponds to the space of
classical solutions. This means in particular that a configuration
must be extendible to a classical solution. In the spacelike
hyperplane case this is trivial. Any (reasonable) scalar function in
space extends to a solution in space-time. For a timelike hyperplane
this is not so. We need to restrict to configurations that do
extend to classical solutions, calling them \emph{physical}
configurations.

Next, we
define ``evolution'' in an interval $[x_1,x_1']$ by the path
integral over the space-time region defined by this interval, in
analogy to (\ref{eq:tevol}). To evaluate the path integral we proceed
in a manner analogous to the spacelike case
\footnote{Explicit computation shows that the
  restriction to physical configurations does not break the
  composition property of the path integral, as it may appear to.}.
The role of positive and negative energy contributions is now played
by contributions with positive and negative sign of the 
momentum component in the $x_1$-direction.
We arrive at the field propagator
\begin{equation}
 Z(\varphi,\varphi')=N \exp\left(-\frac{1}{2}\int\xd t\,\xd^2\tilde{x}
 \begin{pmatrix}\varphi & \varphi' \end{pmatrix} W
 \begin{pmatrix}\varphi \\ \varphi' \end{pmatrix}\right) .
\label{eq:propsbdy}
\end{equation}
The operator-valued matrix $W$ is given by
\[
 W=\frac{\im\pop_1}{\sin\pop_1\Delta}
 \begin{pmatrix}\cos\pop_1\Delta & -1 \\
  -1 & \cos\pop_1\Delta \end{pmatrix} ,
\]
where $\Delta=|x_1'-x_1|$ and
$\pop_1=\sqrt{-\partial_0^2+\sum_{j\ge 2} \partial_j^2-m^2}$.

We turn to the computation of
the vacuum state. As already mentioned, we take as the characterizing
condition the invariance under propagation between parallel
hyperplanes. Making the ansatz (\ref{eq:vacansatz}) with respect to
the $(t,x_2,x_3)$-hyperplane yields the condition
$A^2=\pop_1^2$, similarly to the spacelike case. We set $A=\pop_1$
and justify our choice of sign in a moment.

Given the candidate vacuum state we have defined we proceed to
generalize it to arbitrary timelike hyperplanes by using Lorentz
transformations. Without loss of generality we can restrict again to
Lorentz boosts in the $(t,x_1)$-plane. Thus, consider the boost with
parameter $\gamma$ transforming coordinates via
(\ref{eq:boost}), recall Figure~\ref{fig:lorentz}. Again we
express everything in terms of Euclidean coordinates and angles
according to Figure~\ref{fig:hyperplane}. Since now the boosted
hyperplane is spanned by $(\that,x_2,x_3)$ we find that the angle
$\alpha$ in Figure~\ref{fig:hyperplane} corresponds to the angle
$\tilde{\alpha}$ in Figure~\ref{fig:lorentz}. The coordinate $s$ is
related to $\that$ by the time dilation factor $\rho$ (the same $\rho$
as above) via
$\that=\rho s$. The relation between $\alpha$ and $\rho$ now takes the
form $\cos 2\alpha=-\rho^2$.
Transforming the operator $\pop_1$ as well as the
integration measure in the expression for the vacuum functional,
we obtain precisely formula (\ref{eq:vacgeneral}) again. However,
this time the
range for $\alpha$ is $\pi/4<\alpha\leq\pi/2$. The agreement
with the spacelike case in the limit $\alpha\to \pi/4$ fixes the sign
ambiguity in the choice of $A$.

\section{Particles on timelike hyperplanes}

While the expression for a vacuum state found above is
very suggestive, a convincing justification that
states on timelike hyperplanes are sensible requires us to consider
particle states. We may expect crucial differences 
between such particle states and those on spacelike
hypersurfaces. It will be sufficient to consider particle states
on the hyperplane spanned by $(t,x_2,x_3)$.

Since a state is a function on physical configurations on the
hyperplane, a basis of one-particle states may be characterized by the
Fourier modes in this hyperplane. In the standard spacelike case
these are labeled by 3-momentum. In the present timelike case these
are labeled by the energy and the momentum in the $x_2$- and
$x_3$-directions. The restriction of the energy to satisfy
$E^2\ge\tilde{p}^2+m^2$ corresponds precisely to the
restriction for configurations to physical ones.

Comparison with the spacelike situation (\ref{eq:pstate}) suggests
that a one-particle state should be described by the functional
\begin{equation}
\Psi_{E,\tilde{p}}^{\pm}(\varphi)=\varphif^{\pm}(E,\tilde{p})
 \Psi_0(\varphi) ,
\label{eq:tstate}
\end{equation}
where now $\varphif$ is the Fourier transform in the hyperplane,
\[
\varphif^{\pm}(E,\tilde{p})=2p_1 \int \xd t\,\xd^2 \tilde{x}\,
 e^{\pm\im  (E t - \tilde{p} \tilde{x})} \varphi(t,\tilde{x}) .
\]
We use here the convention that $E\ge 0$ and the actual sign of the
energy is encoded in the extra index $\pm$. Indeed, one can check that
the state (\ref{eq:tstate}) is an eigenstate under propagation between
parallel hyperplanes. Its eigenvalue is $e^{\im p_1\Delta}$ where
$\Delta=|x_1'-x_1|$ and $p_1$ is the positive square root of
$E^2-\tilde{p}^2-m^2$.

What is the interpretation of such states? In the spacelike case
causality prescribes that a state must be purely incoming or
outgoing depending on whether it lies at the beginning or the end of
a (time-)evolution process. An analogue of this does not hold
in the timelike
case. Any individual particle might now be either incoming or
outgoing. This choice is reflected in (\ref{eq:tstate}) by the index
$\pm$, representing the sign of the energy value. To agree with the
spacelike case (concerning the sign of the momentum components $p_2$
and $p_3$) we identify $\Psi^-$ as an in-particle state and $\Psi^+$
as an out-particle state.

In spite of this apparent extra choice in the timelike case, the
degrees of freedom of a particle are identical to those in the
spacelike case.
The reason is that the ``missing'' sign of the momentum
component $p_1$ in the timelike case must be correlated with the sign
of the energy. Namely, the momentum of an in-particle needs to point
from the hyperplane into the propagation region and that of an
out-particle outward. Thus, in both the space- and the timelike case
we may characterize a particle by its 3-momentum.

Complex conjugation converts an in-particle state
$\Psi_{E,\tilde{p}}^-$ to an out-particle state
$\Psi_{E,\tilde{p}}^+$ and vice versa. Moreover, if we put the
complex conjugate of a state on the opposite hyperplane (i.e., on the
other side of the evolution region), then it describes the \emph{same} 
state in terms of the particle 3-momentum. This also parallels the
situation in the spacelike case, although it is more implicit there.
Namely, consider a transition amplitude from a 1-particle state with
momentum $p$ to a 1-particle state of momentum $p'$,
\begin{equation}
 \langle\Psi_{p'}|U(\Delta)|\Psi_p\rangle=
 \int \xD \varphi\xD \varphi'\, \Psi_p(\varphi)
 \overline{\Psi_{p'}(\varphi')}
 Z(\varphi,\varphi') .
\label{eq:tampl}
\end{equation}
The usual interpretation of course is that the complex conjugation
comes from the inner product, as we are pairing a bra- with a
ket-state. However, we could equally well say that a change of
orientation of the hyperplane on which the state lives requires a
complex conjugation to describe the original state. That is, while on
the in-oriented hyperplane a 1-particle state of momentum $p'$ is
described by the wave functional $\Psi_p$, on the out-oriented
hyperplane it is described by the wave functional
$\overline{\Psi_{p'}}=\Psi_{-p'}$.
While this difference of interpretation seems of little consequence in
the spacelike case, it becomes significant in the timelike
case. Indeed, the bra-ket notation is inherently linked to an
orientation of the time-axis chosen from the outset, i.e., we can
distinguish between ``earlier'' and ``later''. In space, on the other
hand,
we can continuously rotate any orientation into its opposite, so fixing
an orientation from the outset makes no sense.

We turn to the dynamics in the form of transition amplitudes.
Mindful of the inadequacy of the bra-ket notation we denote
an amplitude between states $\Psi$ and $\Psi'$ by
$[\Psi,\Psi']$. 
States are understood as specified
with respect
not only to the hyperplane they live in, but also with respect to its
orientation relative to the propagation region. Thus, no explicit
complex conjugation as in (\ref{eq:tampl}) appears. Concretely, a
transition amplitude between states $\Psi$ and $\Psi'$ on
parallel hyperplanes located at $x_1$ and $x_1'$ respectively 
takes the form
\begin{equation*}
 [\Psi,\Psi']=
 \int \xD \varphi\xD \varphi'\, \Psi(\varphi)\Psi'(\varphi')
 Z(\varphi,\varphi') .
\end{equation*}
Here the integrals are over physical configurations $\varphi,\varphi'$
on the hyperplanes and $Z(\varphi,\varphi')$ is the propagator
(\ref{eq:propsbdy}).

Inserting an in-particle of momentum $p$ on the hyperplane at
$x_1$ and an out-particle of momentum $p'$ at $x_1'$ we find the
amplitude to be
\[
2 p_1 (2\pi)^3 \delta(E-E')
\delta^2(\tilde{p}-\tilde{p}')e^{\im p_1\Delta}.
\]
On the other hand, the transition amplitude between in-particles on
both hyperplanes or out-particles on both hyperplanes is zero.
Both results are exactly what should be expected. The probability
(and hence the amplitude) of having two particles go into the
propagation region and none coming out should be zero, and vice versa.
In contrast, having one particle coming in and one going out
should mean in the non-interacting theory we consider that they are
the ``same'' particle, hence the delta functions. The latter situation
parallels the spacelike case.

We finally turn to multi-particle states. As in the spacelike case, an
$n$-particle wave functional is a certain polynomial of degree $n$ in
functionals
$\varphif^{\pm}(E,\tilde{p})$ and delta functions, multiplied with the
vacuum functional. Indeed, the structure of the polynomial is exactly
the same as in the spacelike case (see \cite{Jak:schroedinger} for the
latter). One only has to take into account
that the sign in the delta functions relating to one in- and one
out-particle will be reversed due to our convention of
``externalizing'' the sign of the energy. Furthermore, delta functions
relating to two in- or two out-particles may be dropped as they can
never be satisfied.

\section{A word on unitarity}

The failure of the bra-ket notation in the timelike case indicates
that the role and interpretation of the inner product has to be
considered carefully.
Usually, an inner product and associated probability interpretation is
implemented early on in the construction of a quantum mechanical model,
before the dynamics. Consistency then requires that the dynamics be
compatible with the inner product, i.e., be unitary.

The present context suggests a different route.
We start by defining state spaces associated to hyperplanes,
and transition amplitudes. States on the same hyperplane, but with
different orientation belong a priori to different spaces. However,
physically there is a correspondence between such states, i.e., we
know whether a given in-particle is the ``same'' (has the same 3-momentum)
as a given out-particle. We saw that this identification is given
by the complex conjugation of the wave functional.

We can now
demand that evolving a given state for 0 time duration (or space
distance) yields the same state with probability one. The resulting
expression yields the normalization condition for states,
$ \int\xD \varphi\, \Psi(\varphi)\overline{\Psi(\varphi)} = 1$,
where the integral is over \emph{physical} configurations only.
Indeed, this gives rise to an inner product and with it the possibility
of evaluating the overlap of states etc.
Unitarity is now hidden in the requirement that identification
between states on opposite hyperplanes be preserved under
propagation. Thus, we recover an essentially conventional
probability interpretation.
However, it is important to recognize that
probabilities are in general \emph{conditional} probabilities
implicating the whole measurement process.
More precisely, amplitudes give rise to probabilities for certain
particles on \emph{both} hyperplanes to be observed, conditional on
certain other particles on \emph{both} hyperplanes to be
present (prepared).

\acknowledgments

I thank the Perimeter Institute in Waterloo, Canada, for hospitality,
where part of this research was carried out.

\bibliography{stdrefs}

\end{document}